# Experimental Signatures of a New Channel of the Deuteron-Deuteron Reaction at Very Low Energy

R. Dubey,[1,*] K. Czerski,[1] Gokul Das H.,[1] A. Kowalska,[2] N. Targosz-Sleczka,[1] M. Kaczmarski,[1] and M. Valat[1]

[1]*Institute of Physics, University of Szczecin, 70-451 Szczecin, Poland*
[2]*Institute of Mathematics, Physics and Chemistry, Maritime University of Szczecin, 70-500 Szczecin, Poland*

The discovery of a new, $e^+e^-$ reaction channel in deuteron-deuteron (DD) fusion at very low energies might have major implications for understanding primordial and stellar nucleosynthesis, where electron-positron reaction channels are typically not considered. It could also enable research on metal hydride fusion, potentially paving the way for the design and construction of next-generation fusion energy sources. Following the first experimental indications of electron emission, we present here an extensive experimental study confirming emission of high-energy electrons from DD reaction at very low energy. A simultaneous use of Si charged particle detectors of different thicknesses and large-volume NaI(Tl) and HPGe detectors has allowed the determination of the branching ratios between emitted protons, neutrons, and $e^+e^-$ pairs for deuteron energies down to 5 keV. The high-energy positrons could be unambiguously detected by their bremsstrahlung spectra and annihilation radiation, supported by the Geant4 Monte Carlo simulations. The theoretical calculations, based on a destructive interference between the threshold resonance and the known broad resonance in $^4$He, agree very well with experimentally observed increase of branching ratios for lowering projectile energies. The partial width of the threshold resonance for the internal $e^+e^-$ pair creation should be at least 10 times larger than that of the proton channel.

## I. INTRODUCTION

Primordial nucleosynthesis, which occurred within the first few minutes of the Universe's existence, produced the lightest elements. Deuterium, in particular, serves as a sensitive probe of early cosmic conditions, as its abundance is strongly linked to baryon density and the neutrino population [1–3]. While precise deuterium measurements exist, theoretical predictions of its formation have been inhibited by uncertainties in the deuterium-burning reaction rate. Likewise, the formation of helium-3 and helium-4 nuclei via deuteron-deuteron (DD) and deuteron-proton (DP) fusion reactions under specific plasma conditions is crucial for understanding stellar nucleosynthesis, especially in objects such as brown, red, and white dwarfs as giant planets [4]. In these kinds of objects, the electron screening effect, shielding the Coulomb barrier between reacting nuclei, leads to an increase of nuclear reaction rates by many orders of magnitude [5].

In recent years, another intriguing effect in the DD reactions at very low energies has been observed: indications for the existence of a $0^+$ DD threshold resonance and its decay via internal pair creation (IPC) [6,7]. According to our first measurements, this channel is expected to dominate the reaction cross section below a deuteron energy of 5 keV. In this energy range, screening of the Coulomb barrier by quasifree electrons in a deuterated metallic target can enhance the DD fusion cross sections by more than one order of magnitude, depending on the electronic and crystal lattice properties of the hosting target material [7]. Both effects are also of large importance for the design and construction of next-generation clean energy sources [7].

Theoretically, DD threshold resonance occurs due to weak coupling between the $2 + 2$ and $3 + 1$ cluster states in the compound $^4$He nucleus [6], as originally proposed in 1972 [8] and explained through multichannel $R$-matrix parametrization [9]. This theory has recently been supported by experimental results [10,11]. However, this effect cannot be correctly reproduced in *ab initio* calculations [12–15] which predict a significant contribution of the DD continuum to the first excited state $0^+$ in $^4$He, being a cluster state of proton and triton, located about 3.6 MeV below the DD threshold [6]. In the case of strong coupling, the threshold DD resonance would merge with the first excited state as demonstrated in Refs. [16,17], and the DD reaction would favor neutron and proton channels compared to the 7 orders of magnitude weaker $\gamma$ emission.

[*]Contact author: rakesh.dubey@usz.edu.pl

Experimentally, measurement of the DD reactions at extremely low energies is very challenging not only because of the drop in the fusion cross sections but also because of atomic effects which can contribute to the yields of measured reaction rates. First, the enhanced electron screening effect, which is responsible for reducing the height of the Coulomb barrier between the reacting nuclei, depends on the type of the crystal lattice structure as well as the occurrence of crystal lattice defects in the host metallic target [7,18,19]. The electron screening effect leads to an exponential increase in the reaction probability in comparison to the case of bare nuclei.

However, all these effects can be comprehensively examined using an ultrahigh vacuum low-energy accelerator facility [20]. Moreover, to reduce systematic experimental uncertainties, the electron-to-proton branching ratios (BRs) were measured directly by applying single Si detector technique [10,21]. These measurements, together with theoretical calculations and Geant4 Monte Carlo (MC) simulations [22], supported the preliminary idea of DD threshold resonance and its decay via IPC [10,21]. However, apart from relatively poor event statistics, no experimental arguments for precise energy determination of electrons and positrons, or indication of high-energy positron emission, could be provided.

Therefore, in this work, a much more reliable signature of a novel reaction channel in the DD fusion at ultralow energies is presented. Following the initial observation, a two-year experimental campaign to get better insights into the DD threshold resonance mechanism was performed. To meticulously examine this new phenomenon by measuring the electron/proton BR, a long-lasting series of experiments was employed. In these experiments, various charged particle detector setups (Si detectors and Al absorption foils with varying thickness) at very low deuteron beam energies ranging between 5 and 20 keV were used.

Additionally, for the first time, we measured bremsstrahlung and annihilation radiation resulting from the decay of DD threshold resonance using large-volume NaI(Tl) and HPGe detectors. Furthermore, the theoretical reaction model was refined, resulting in a substantially increased estimate of the partial resonance width associated with the IPC decay channel of the DD threshold resonance.

## II. EXPERIMENTAL PROCEDURE

The experiments have been performed at the eLBRUS Ultra High Vacuum Accelerator (UHVA) facility of the University of Szczecin, Poland. A schematic diagram of the experimental setup consisting of a silicon detector, aluminum absorber foils in front of the Si detector, a target inside the target chamber, and a large-volume NaI or HPGe detector placed outside the chamber and corresponding geometry simulated with Geant4 and visualized with OpenGL is shown in Fig. 1.

The UHVA facility combines a very good vacuum of $10^{-10}$ mbar in the target chamber with a high current ion beam up to 1 mA on the target [20]. As cross sections of nuclear reactions proceeding far below the Coulomb barrier decrease exponentially with lowering projectile energies, particular attention has been paid to the precise energy definition of the beam provided by the electron cyclotron resonance ion source, being of about 10 eV with a long-term stability of about 5 eV. The ion beam is analyzed magnetically and focused by a series of electric lenses to form an elliptical beam with a $7 \times 12$ mm$^2$ spot on the deuterium-implanted zirconium (Zr).

Deuterium $D^+$ and $D^{2+}$ ion beams (40–60 μA) current with energy ranging between 5 and 20 keV was used during experiment. The Zr target was tilted at 45° to the beam direction and it was prepared from foils with a thickness of 1 mm and 99.2% purity and cut to dimensions of $1.5 \times 2$ cm$^2$. Zr material was chosen as a target because it is known for its stability, high hydrogen stoichiometry, and the homogeneity that can be achieved. At such low $D^2$ beam energy in metallic environments, electron screening strongly affects DD fusion, and this effect depends on vacancy formation during deuteron irradiation. Even minor impurities of carbon and oxygen could influence the DD fusion process by changing vacancy depth distribution and diffusion [19].

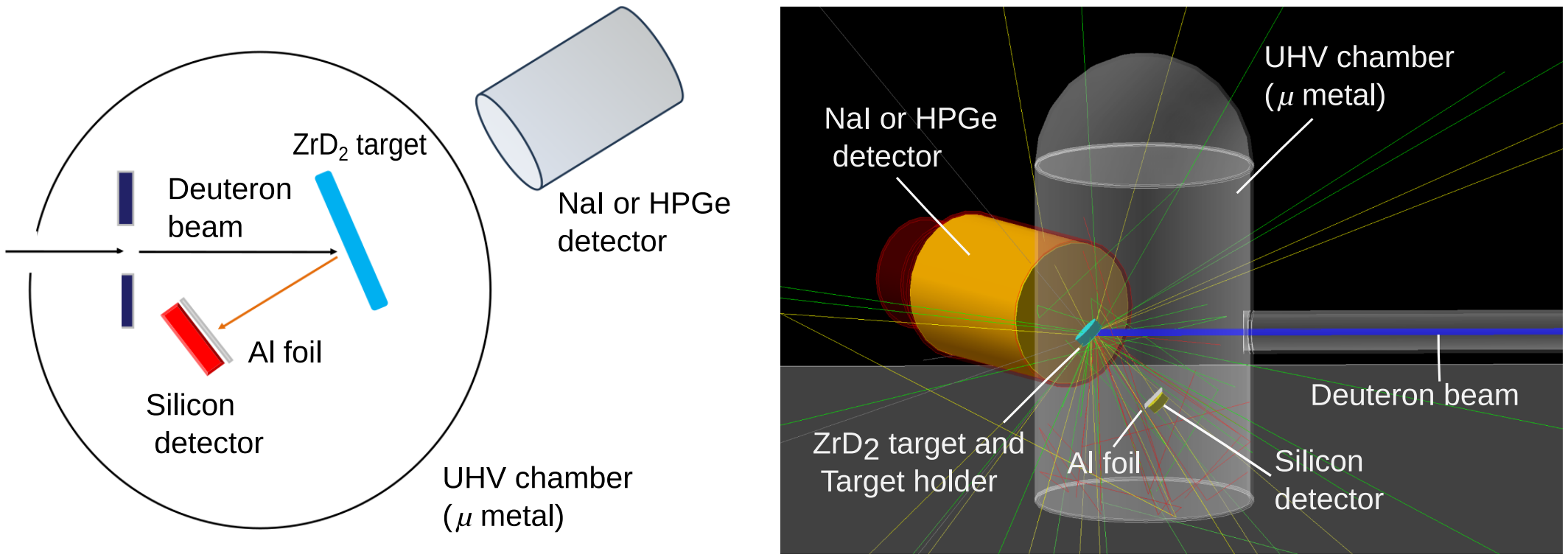

FIG. 1. Schematic view of the experimental setup with one silicon detector. Aluminium absorber foils are placed in front of the Si detector. Large-volume NaI(Tl) and HPGe detectors are placed outside the chamber at 135°. Inside view of detection geometry consisting of a silicon detector, target, and target holder simulated with Geant4 and discussed in detail [21].

To experimentally measure the high-energy IPC and BR of IPC over protons and IPC over neutrons in the DD reaction, we employed three different kinds of detectors. For the most precise measurements, we placed a single surface-barrier Si detector with different thicknesses (1, 2, and 3 mm) at a distance of 7 cm from the target and it was positioned at an angle of 135° with respect to the beam direction. In front of the detectors, Al absorption foils of different thicknesses between 1 and 125 μm were placed to stop elastically scattered deuterons as well as 0.8-MeV $^3$He, 1.02-MeV $^3$H, and 3.02-MeV protons. For comparison, the energy spectrum measured with a very thin Si detector (100 μm thick) and an absorption Al foil (1 μm thick) is presented in Fig. 2.

High-energy 22.82-MeV IPC will be partially absorbed by Si detectors (with thicknesses of 1–3 mm) resulting in characteristic absorption energy lines in the detectors. Since the stopping power of electrons with energies ranging between 1 and 20 MeV is almost constant, the energy absorbed in the detectors varied only slightly and the maximum of the measured energy spectrum was proportional to the detector thickness. Therefore, the continuous $e^+e^-$ energy spectrum could be observed as a relatively narrow absorption bump, strongly improving the effect-to-background ratio. However, to accurately determine the strength of the IPC transition using a single Si detector, MC Geant4 simulations were necessary to account for the effects of elastically scattered protons and electrons within the target, target holder, and protective foils on the measured spectrum [21].

The detection method using a single 1–3-mm-thick Si detector cannot present the full energy spectrum of emitted high-energy 22.82-MeV $e^+e^-$ pairs; we can only obtain the partially deposited energy in this detector. Therefore, it is difficult to infer exactly the energy of $e^+e^-$ pairs, especially when they have greater than 5-MeV energy [21]. Hence, to further confirm the signature of 22.82-MeV $e^+e^-$ pair emission from DD fusion, we have additionally placed a large-volume cylindrical NaI(Tl) detector (20 cm diameter, 30 cm length) coupled to four separate photomultipliers, which is able to detect the high-energy bremsstrahlung related to the 23.84-MeV IPC transition. The NaI(Tl) detector was placed outside the target chamber at a distance of 32 cm from the target sample at the 135° angle with respect to the beam direction.

As shown in Fig. 3, the energy efficiency of the detector was determined using Geant4 MC simulations and fitted to the experimental data obtained for calibration sources: $^{137}$Cs, $^{60}$Co, and $^{22}$Na. The efficiency curve was extended to high energies using Geant4 MC simulation and used to calculate a proper energy spectrum of the bremsstrahlung induced by the internal $e^+e^-$ pair creation and to estimate the excess of positron annihilation events based on the count rate of the 511-keV annihilation line. More details about detector performance parameters such as calibration, detector resolution, response function for high-energy $e^+e^-$ pairs using Geant4 MC simulations are presented in a recently accepted work [23] and in the Appendix.

The experimental $\gamma$ spectrum of the NaI detector obtained during irradiation of a deuterated Zr target ($ZrD_2$) with a 10-keV deuteron beam is shown in Fig. 4 along with foreground and background (when the beam stopped at the Faraday cup). The background at the Faraday cup includes all kinds of accelerator-induced events before the beam hits the target and background measurements obtained when the accelerator was fully turned off. The background measured in the same energy region (especially 8–23 MeV) with the ion beam on the Faraday cup is significantly higher, primarily due to electromagnetic noise induced by the accelerator; however, its shape remains nearly energy independent.

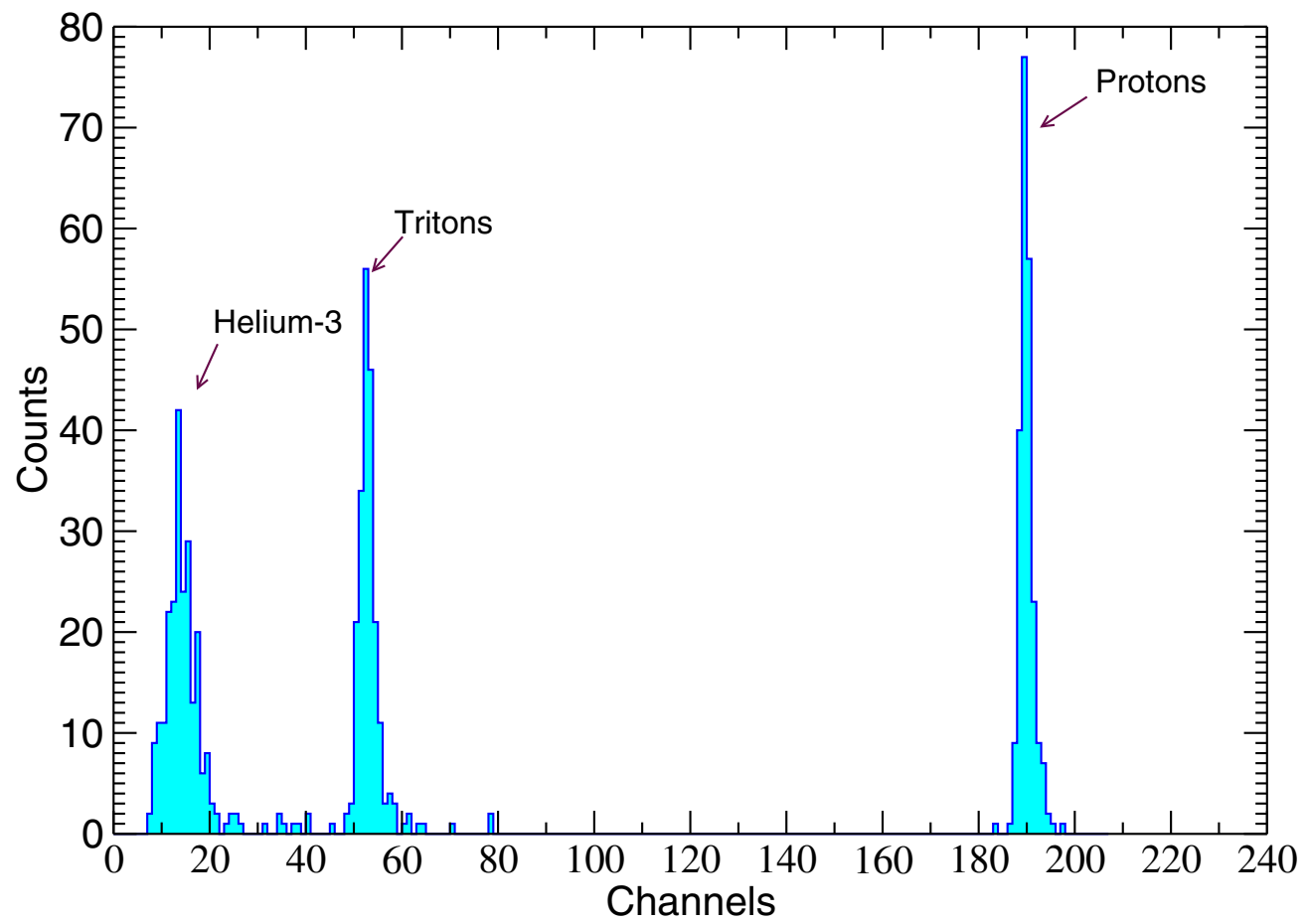

FIG. 2. Energy spectrum measured with 100-μm-thick Si detector and 1-μm-thick Al absorption foil at 10 keV $D_2$ beam. Because of the very small thickness of the detector, no electron or positron bump could be observed.

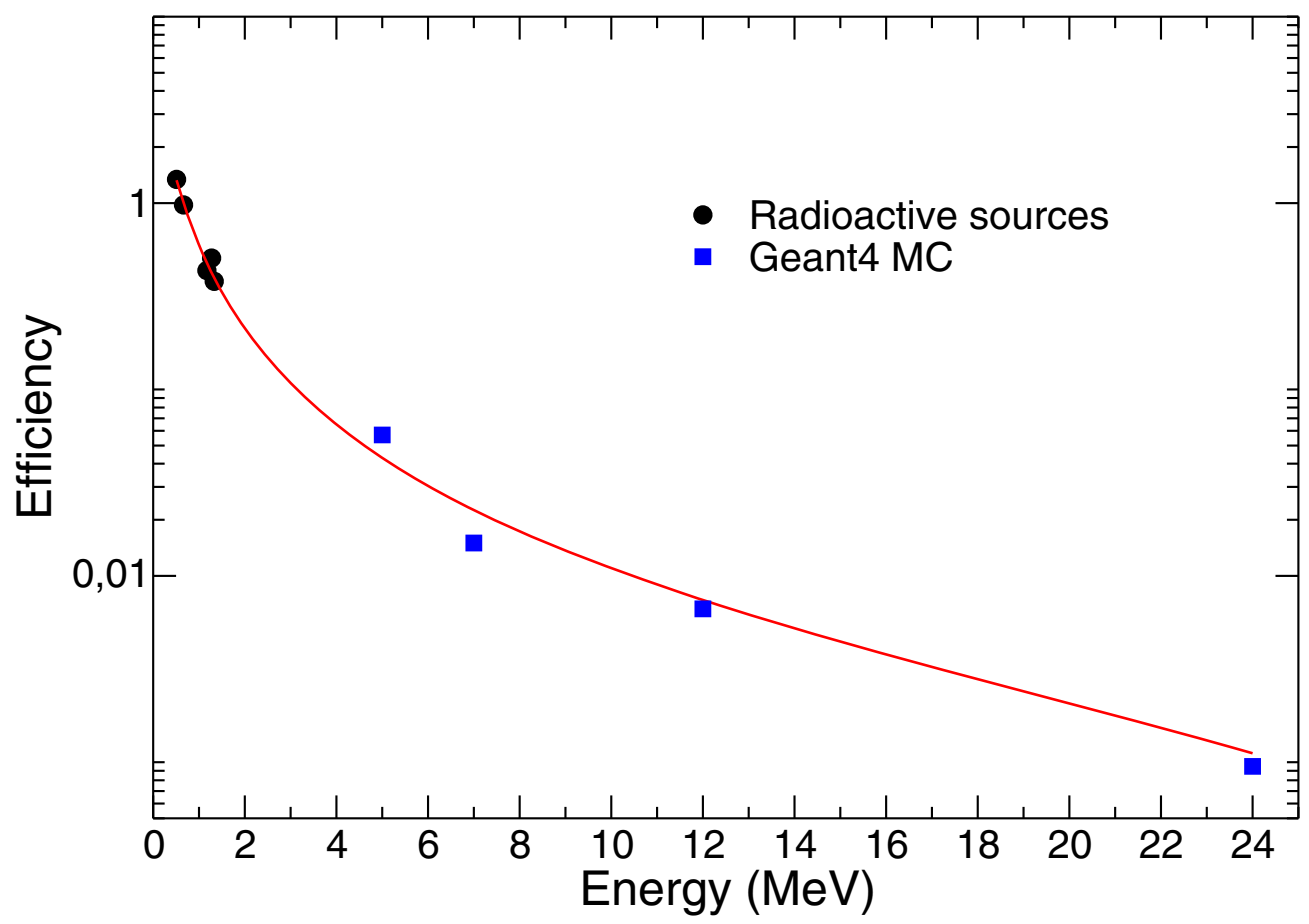

FIG. 3. Efficiency curve for a large-volume NaI(Tl) detector (20 cm radius, 30 cm length) with experimental data obtained for different radioactive sources. High-energy efficiency was estimated using Geant4 MC simulation, which was verified by existing calibration for the same size detector [24,25].

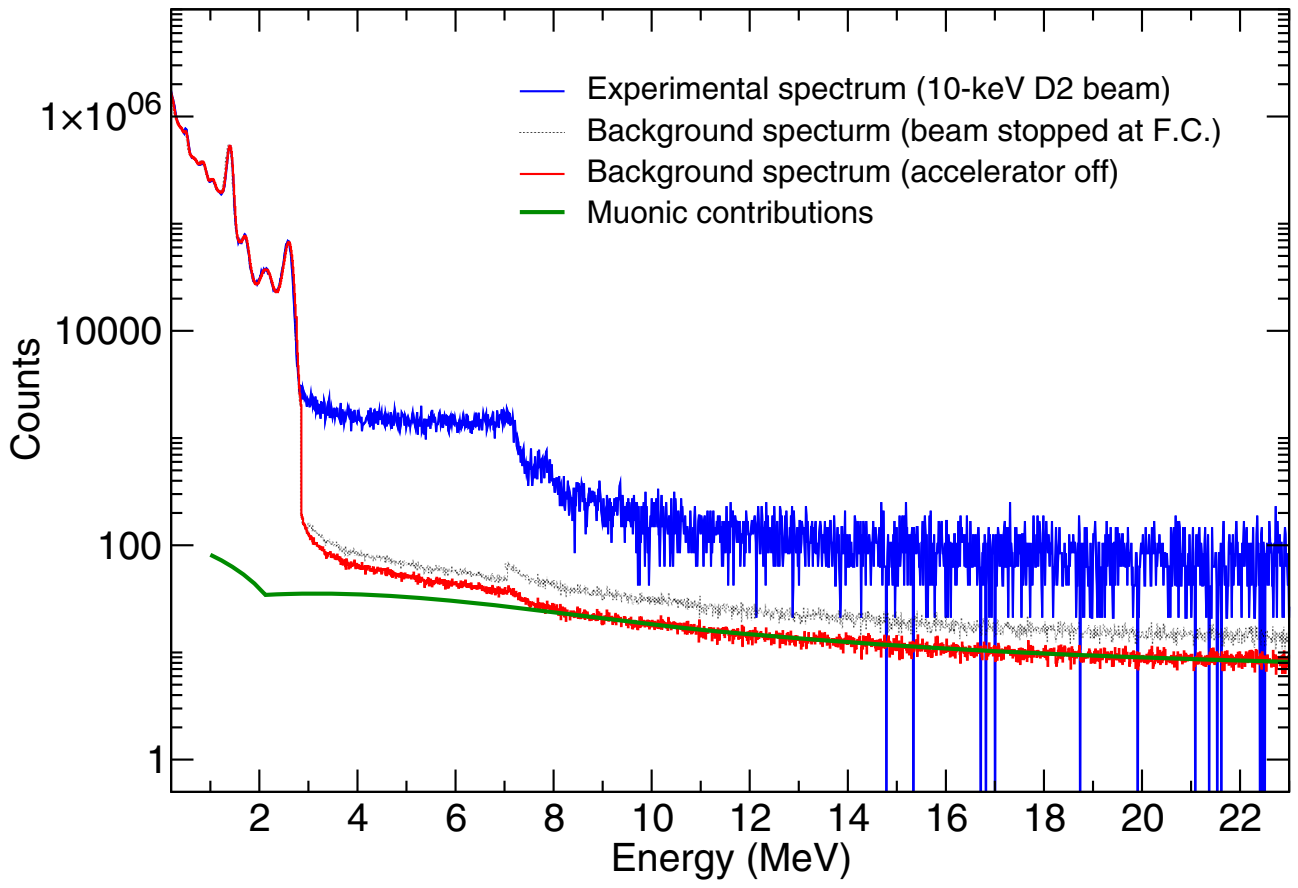

FIG. 4. Experimental $\gamma$ spectrum (blue curve) measured with a 10-keV deuteron beam on the $ZrD_2$ target, compared to background measurements obtained when the accelerator was fully turned off (red curve) and with the beam stopped at the Faraday cup (black curve). The green solid line shows the estimated cosmic muonic contributions [24,26,27].

Further, the low-energy $\gamma$ spectrum has also been studied with an HPGe detector, taking advantage of its much better energy resolution compared to the NaI detector. 511-keV peak strength can be more clearly resolved for the IPC-induced weak effect compared to the natural background in the HPGe detector spectrum than in the NaI detector. The large-volume HPGe detector has a 58-mm crystal diameter and 66-mm length, providing 35% relative efficiency and 2-keV energy resolution (FWHM) in the experimental setup.

## III. RESULTS

### A. Charge particle spectra

Figures 5(a)–5(c) display the charge particle spectra obtained after background subtraction from 2- and 3-mm-thick Si detectors following DD reactions at 10-keV deuteron beam energy. A prominent peak at 0.7 MeV and at around 1 MeV, measured with 2- and 3-mm Si detectors, respectively, corresponds to the $e^+e^-$ IPC according to the Geant4 calculations (red curve). At higher energies, 3-MeV protons are also visible in Figs. 5(a) and 5(b). Their measured energy peak positions of 1.8 and 2.2 MeV are due to energy absorption in the 46- and 36-μm-thick Al foils, respectively, placed in front of the detectors. Such thicknesses of the Al foils were enough to fully absorb 1.02-MeV tritons and 0.8-MeV $^3$He particles, so they are not visible in the spectra. In the case of the thickest, 125-μm Al foil

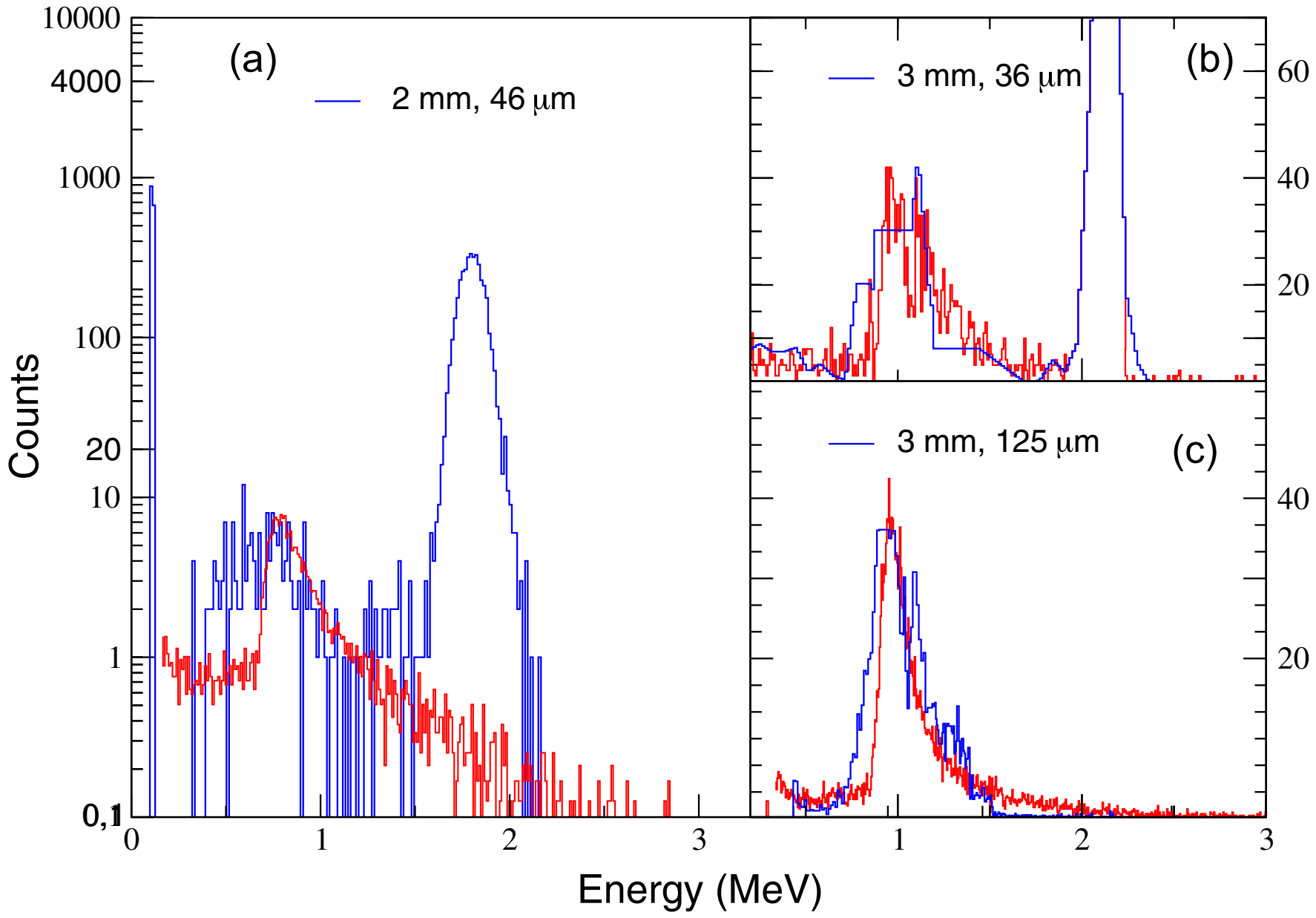

FIG. 5. The experimental energy spectrum (in blue) measured at deuteron energy of 10 keV using (a) 2-mm-thick Si detector with 46-μm absorption Al foil, (b) 3-mm-thick Si detector with 36-μm absorption Al foil, and (c) 3-mm-thick Si detector with 125-μm absorption Al foil. In the last case, the protons are fully absorbed in the Al foil. The Geant4 calculated spectrum accounting for $e^+/e^-$ IPC is presented in red.

[Fig. 5(c)], the proton line also disappears. In this configuration, only a distinct $e^+/e^-$ peak persists around 980 keV due to their much lower stopping power values.

As discussed in Refs. [10,21], the position of the $e^+/e^-$ peak corresponds to the partial energy deposition of high-energy electrons and positrons within Si detectors, which are not thick enough to absorb their full energy. As the stopping power functions of electrons and positrons are very similar and change only slightly in the energy range of 1–20 MeV, the peak position corresponds to the thickness of the detector. Detailed calculations of the electron and positron spectra were performed, taking into account both the continuous $e^+e^-$ energy distributions arising from the IPC $0^+ \to 0^+$ transition to the ground $^4$He state, as well as the comprehensive Geant4 MC simulations of the experimental setup [21]. The slight discrepancy observed in the peak position of the simulated versus measured electron spectrum can be attributed to nonuniformities in the depletion region of the thicker Si surface barrier detectors. To account for this effect, a correction of approximately 10% for 2- and 3-mm-thick detectors was incorporated into the simulations. It is important to emphasize that the $e^+e^-$ peak could not be observed in previous measurements performed by many authors since they used thinner Si detectors (of the order of 100 μm) to improve the signal-to-noise ratio (see Fig. 2) [7]. In that case, the $e^+e^-$ absorption peak would lie at energy of about 30 keV, being below the noise threshold of Si detectors, disabling its observation.

The CRY+Geant4 combination has so far proven to be the most suitable for modeling cosmic rays in silicon (Si) material [28,29]. It has been observed that $\mu^\pm$ and $e^\pm$ dominate the interactions in terms of particle counts. The spectra for both charge states of muons ($\mu^\pm$) and pions ($\pi^\pm$) are very similar, while electrons ($e^-$) outnumber positrons ($e^+$) over most of the spectrum. High-energy electrons and positrons, as well as low-energy muon events, can produce overlapping partial energy peaks, even in thick Si detectors.

The strength of the partial energy peak induced by IPC is compared to that induced by cosmic particles ($e^\pm$, $\mu^\pm$), and is found to be nearly 40 times stronger for a 2-mm-thick Si detector, and 25 times stronger for a 3-mm-thick Si detector for 10-keV deuteron beam energy.

### B. NaI(Tl) photon spectra

All components of the NaI(Tl) photon energy spectra compared to the Geant4 MC simulation calculations are shown in detail in Figs. 6(a) and 6(b). The neutron response function of all elements of the experimental setup has been simulated with the Geant4 MC code, adopting the ENDF-VII.0 model as a dataset of the cross section for neutrons [30]. The BROND [31], JEFF [32], and JENDL [33] libraries, which are often used in simulations, show slight discrepancies with our experimental data. The simulated spectra have been smeared with a Gaussian function to reproduce the measured resolution of the NaI crystal, enabling the determination of the number of emitted neutrons from the $^2\mathrm{H}(d,n)^3\mathrm{He}$ reaction at two deuteron energies: 10 and 20 keV. The detailed procedure for unfolding the NaI spectrum is presented in the Appendix. The validity of the performed MC simulations was confirmed with the help of existing works for similar sizes of NaI(Tl) detectors [25].

As depicted in Figs. 6(a) and 6(b), the neutron-induced spectrum, showing several well-resolved discrete transitions, could be easily identified. In the low-energy region (0–3 MeV), the $\gamma$ spectra are dominated by natural radioactivity of $^{40}$K (1462-keV line) and $^{208}$Tl (2615-keV line). In the intermediate energy region 3–10 MeV, a strong contribution induced by the 2.5-MeV neutrons from the $^2\mathrm{H}(d,n)^3\mathrm{He}$ reaction is visible. The photon peaks (at 1.37, 2.75, 4.5, 5.7, and 6.4 MeV) induced by the $^{23}\mathrm{Na}(n,\gamma)^{24}\mathrm{Na}$ reaction on the NaI(Tl) detector are easy to identify, as well as the line at 6.9 MeV, resulting from $^{127}\mathrm{I}(n,\gamma)^{128}\mathrm{I}$ [34,35]. The peaks attributed to the $\mu$-metal (75% nickel, 20% iron) chamber with high ($n$, $\gamma$) cross sections of $^{56}$Fe (92% natural abundance) at 7.645 MeV and other peaks between 8 and 10 MeV could be recognized. Overall, the major contributions of neutron-induced photons in the measured spectra are dominant below 10 MeV, with almost negligible effects of neutron-induced reactions beyond this energy region. At an energy of about 9 MeV of $\gamma$ photons, a clear transition point can be observed, above which the IPC-induced bremsstrahlung radiation begins to dominate.

For confirming the signature of bremsstrahlung produced by high-energy 22.82-MeV IPC $e^+e^-$ pairs, statistics of count rate in the high-energy region (10–23 MeV) play a very important role. However, the count rate obtained in this energy region is significantly influenced by high-energy cosmic events. To reduce statistical and systematic uncertainties, the photon spectrum in this region was first summed over more pulse-height channels [black step curve in Fig. 6(b)] and then, after taking into account the background contribution, the amplitude of the bremsstrahlung spectrum could be fitted. In this energy region, integrated experimental counts after normalized background subtraction nearly match the IPC bremsstrahlung simulation. This exhibit improved agreement with simulation, suggesting that IPC bremsstrahlung processes are well modeled. Raw data are presented in Table II of the Appendix.

The strength of the bremsstrahlung contribution and the neutron-induced part of the $\gamma$ spectrum corrected by the NaI (Tl) detector efficiency have been used to estimate the BR between $e^+e^-$ pair creation and the number of emitted neutrons and protons measured distinctly by the Si detector (see the Appendix and Table I).

### C. Positron annihilation spectra

In the case of the internal $e^+e^-$ pair emission, an excess of counts in the 511-keV annihilation line should be observed by both NaI(Tl) and HPGe detectors under

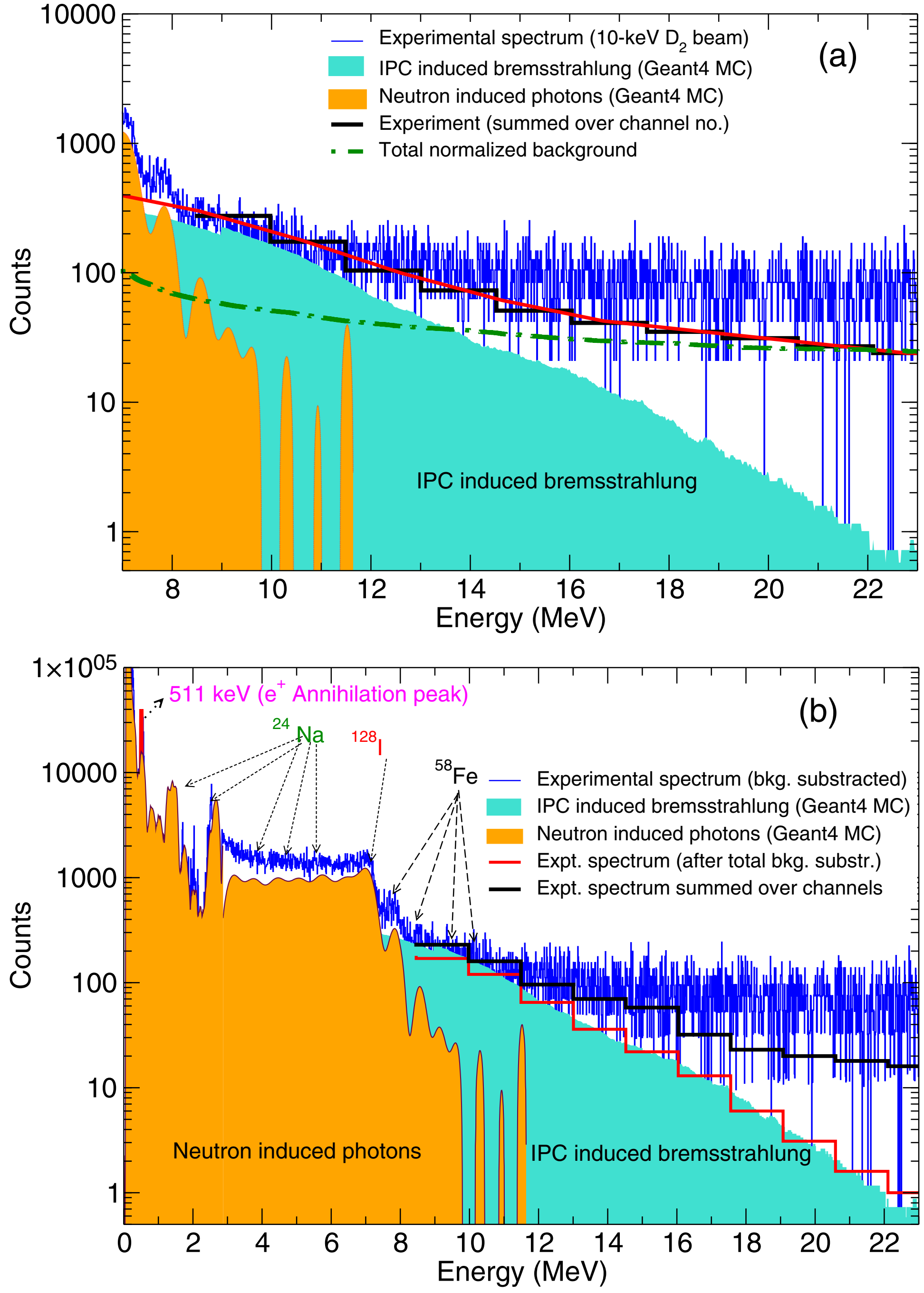

FIG. 6. (a) High photon energy spectrum measured by a large cylindrical NaI(Tl) detector at a deuteron energy of 10 keV (blue curve), experimental spectrum summed over more pulse-height channels (black step curve) to reduce statistical fluctuation, and total normalized background (green dashed curve) are depicted. (b) Full photon energy spectrum measured after subtraction of the natural muonic background (blue curve); the red step curve is the same spectrum obtained after subtraction of the full background. Geant4 Monte Carlo bremsstrahlung spectrum induced by the IPC electrons or positrons (light green color area) has been fitted to the experimental spectrum (red curve). The neutron-induced contribution (yellow color area), calculated using the Geant4 MC simulation, is displayed as well.

deuteron irradiation of the $ZrD_2$ target [36]. The corresponding spectra measured with the 10-keV deuteron beam and the background measurement without running the accelerator are presented in Figs. 7(a) and 7(b). The excess of 511 keV counts is clearly observed, although a large background contribution is also visible. As the strength of the background annihilation line changes in long-term measurements only by about 10%, even the deuteron beam is placed on the Faraday cup in front of the target chamber. Therefore, its origin seems to be natural and not due to the external pair creation of $\gamma$ radiation induced by neutrons from the DD reactions. This is also in agreement with Geant4

TABLE I. Experimentally determined branching ratios between IPC or annihilation radiation (AR) and proton ($p$) or neutron ($n$) channels using different detectors (Si, NaI, HPGe) at 10- and 20-keV deuteron beam energies. Uncertainties result from both experimental statistics and systematic uncertainties (see Appendix). The latter can be reduced by calculating the ratio between values obtained at different beam energies. For comparison, theoretical values are also presented.

| Energy (keV) | $IPC_{Si}/p$ (3-mm Si) | $IPC_{NaI}/n$ (NaI) | $AR_{NaI}/n$ (NaI) | $AR_{HPGe}/n$ (HPGe) | Theory |
|---|---|---|---|---|---|
| 10 | $0.42 \pm 0.05$ | $0.83 \pm 0.30$ | $1.70 \pm 0.50$ | $1.10 \pm 0.40$ | 0.25 |
| 20 | $0.060 \pm 0.012$ | $0.09 \pm 0.04$ | $0.16 \pm 0.04$ | $0.12 \pm 0.04$ | 0.042 |
| 10/20 ratio | $7.0 \pm 1.6$ | $9 \pm 4$ | $10 \pm 3$ | $9 \pm 3$ | 5.8 |

MC simulations which predict 2 orders of magnitude lower annihilation line intensity than the excess effect measured. Taking into account the absolute efficiency of the NaI detector (as shown in Fig. 3), NaI detected the same intensity of high-energy IPC ($e^+e^-$) flux for bremsstrahlung, and 511 keV annihilation was detected by NaI, with reduced intensity of flux in the HPGe detector due to differences in the solid angle and absolute efficiency. The BR of the IPC process relative to the proton or neutron channels can be estimated. In the deuteron beam energy range studied, both reaction channels are assumed to contribute equally. On the other hand, if two different reaction channels can be investigated using the same detector, the systematic uncertainties of their BR related to the detector positions can be neglected. The experimentally determined BRs based on the strength of the annihilation line are comparable to the values obtained using other methods (see Table I) and support the IPC process in the DD fusion at very low energy. Their values are significantly larger for the deuteron energy of 10 keV than for 20 keV in agreement with the results obtained with Si detectors. To estimate the BR while minimizing the propagation of systematic and statistical uncertainties, we employ a ratio method that compares measurements at two different energy points. This approach effectively cancels out common systematic errors. We have implemented this method by calculating the ratio of the BR obtained at 10 keV to that at 20 keV, as presented in the third row of Table I. Details of our analysis procedure for unfolding of NaI(Tl), HPGe photon spectra and corresponding BR calculations are provided in the Appendix.

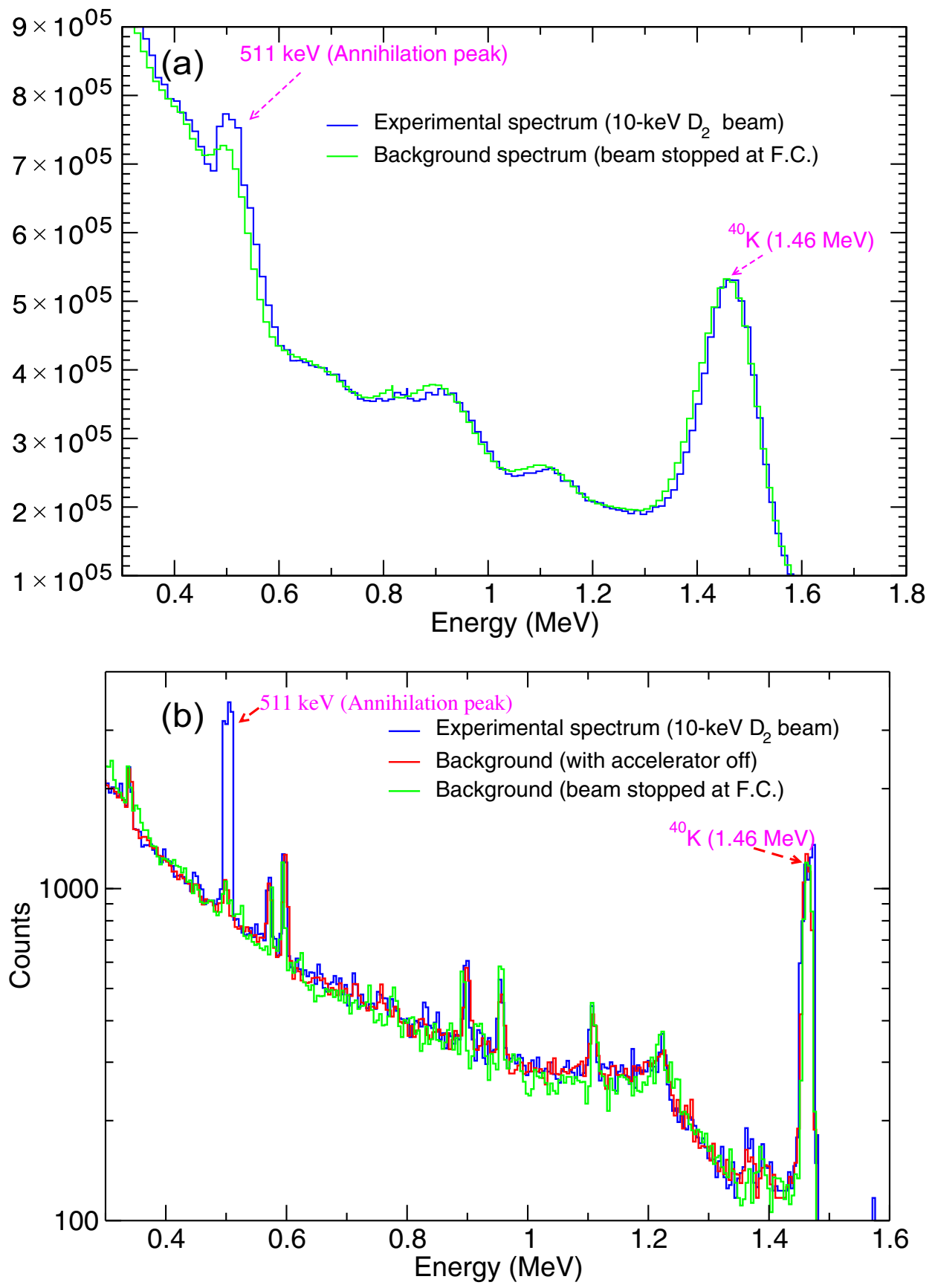

FIG. 7. (a) Low-energy $\gamma$ spectra measured with the NaI(Tl) detector. The red curve corresponds to the background measurement performed with the deuteron beam on the Faraday cup. The blue curve represents the effect obtained during irradiation (10-keV deuteron beam) of the deuterated Zr target. (b) Low-energy $\gamma$ spectra measured with the HPGe detector. The spectrum obtained for the deuterated Zr target irradiated with a 10-keV deuteron beam is compared to the background measurements taken either with the ion beam on the Faraday cup or with the accelerator turned off.

### D. Theoretical calculations

In present work, we theoretically explain our electron observation by considering the existence of the DD threshold resonance in the $^4$He nucleus and its decay via internal $e^+e^-$ pair creation. The experimental electron-proton BR determined for the deuteron energies at 5–20 keV using 1–3-mm-thick Si detectors with different Al absorber foils is shown in Fig. 8. A methodology adopted for the BR calculation using measured thick target yield for proton is well discussed in our recent studies [10,21]. An increase of its value for the lowering deuteron energy related to the DD threshold resonance is clearly visible.

The excitation of this resonance has been previously observed in the $^2$H$(d,p)^3$H reaction on both metallic and gaseous targets [6,7]. From Eq. (1), we can see the proton cross section to be well described as the sum of two components: one corresponding to the well-known overlapping broad resonances in $^4$He and the second describing the narrow-threshold resonance and the interference term between both amplitudes [10]:

$$\sigma_p = \frac{\pi}{k} P(E+U_e)\left[\frac{k}{\pi}\frac{1}{\sqrt{EE_G}} S_p(E) + \frac{2\hbar^2}{\mu a}\frac{\Gamma_p}{E^2} + 2\left(\frac{k}{\pi}\frac{S_p(E)}{3}\frac{1}{\sqrt{EE_G}}\cdot\frac{2\hbar^2}{\mu a}\frac{\Gamma_p}{E^2}\right)^{1/2}\cos(\phi_p^{0+})\right], \quad (1)$$

where $k$, $E$, $E_G$, $a$, $S_p(E)$, $\mu$, $\phi_p^{0+}$, and $\Gamma_p$ denote wave number, deuteron energy in the center-of-mass frame, Gamow energy, channel radius, astrophysical $S$ factor, reduced mass, Coulomb wave function, and proton partial width, respectively. In the present calculation, $a = 7$ fm and the Gamow energy $E_G$ was considered to be 986 keV.

The penetration factor describes the s-wave penetration probability through the Coulomb barrier reduced by the electron screening energy $U_e$, and it can be expressed as

$$P(E+U_e) = \sqrt{\frac{E_G}{E+U_e}} \exp\left(-\sqrt{\frac{E_G}{E+U_e}}\right). \quad (2)$$

The broad resonance component [the first term in brackets in Eq. (1)] can be determined from the known astrophysical $S$ factor for the proton channel equal to about 57 keVb at very low deuteron energies. The second term results from the Breit-Wigner formula for the threshold resonance, assuming that the deuteron partial width takes on the maximum value of the single-particle resonance [7]. This relation could be simplified as the resonance energy is close to zero and therefore much smaller than the deuteron energy. The interference (third) term corresponds to the fact that only one third of the broad resonance transition goes through the $0^+$ compound nucleus state and can interfere with the threshold resonance [6]. The proton partial width of the threshold was experimentally estimated to $\Gamma_p =$ 40 meV and the nuclear phase shift of the broad resonance component to $\varphi_p^{0+} = 135°$.

The cross section for the $e^+e^-$ pair creation can be calculated similarly to the proton channel [see Eq. (1)], leading to an expression where $\Gamma_{IPC}$, $\Gamma_p$, and $S_{IPC}$ should be fitted to the experimentally determined electron-proton BR. Since the penetration factors for both channels are the same, the BR will be independent of the electron screening energy. Differently from the previous calculations [10,11] we consider here that a small IPC component of known broad resonances $S_{IPC}$ exists and can interfere with the IPC contribution of the threshold resonance.

The theoretical calculations for different values of parameters are presented in Fig. 8. The best fit could be achieved for the ratio $\Gamma_{IPC}/\Gamma_p = 14$, $S_{IPC} = 0.25$ keVb, and $\varphi_p^{0+} =$ 170°, assuming $\Gamma_p = 20$ meV. The theoretical curve was especially sensitive to the choice of the nuclear IPC phase shift. Because of its large value, the destructive interference for the IPC channel reaches its almost maximum value.

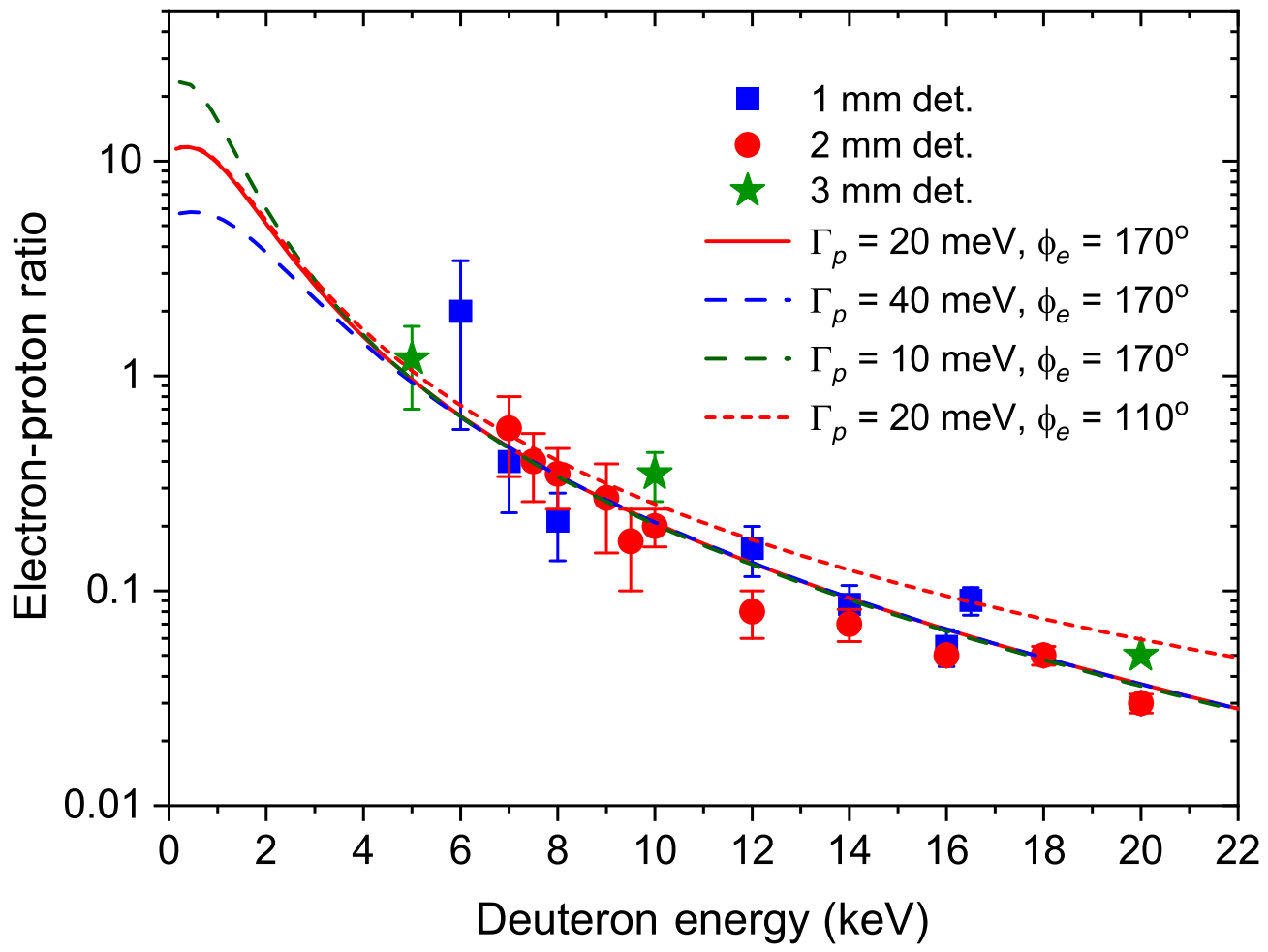

FIG. 8. Experimentally determined IPC over proton BR compared to the theoretical calculations assuming different proton partial width of the DD threshold resonance and the phase shift of the broad resonance contribution.

On the other hand, small changes in the proton partial resonance width do not influence the fit quality significantly within the range of the experimental data. Only differences can be observed at deuteron energies below 2 keV. A small IPC contribution arising from the broad resonances was necessary to obtain a better fit to the new experimental data, which simultaneously leads to an approximately 6 times higher $\Gamma_{IPC}/\Gamma_p$ ratio than determined before.

## IV. DISCUSSION AND CONCLUSIONS

We have presented new measurements of the $\Gamma_{IPC}/\Gamma_p$ BR of DD reactions down to the lowest deuteron energy of 5 keV, using Si detectors of different thicknesses (1–3 mm) and strongly increasing the number of experimental points to reduce statistical uncertainties. For the first time, we have also observed both annihilation and bremsstrahlung radiation with large-volume NaI(Tl) and HPGe detectors, which confirms emission of electrons and positrons with energy up to 23 MeV in DD reaction at very low beam energies. The unambiguous evidence for positron emission has been the observation of an excess on the 511-keV annihilation line.

The analysis of experimental data has been supported with Geant4 MC simulations that allowed the identification of a broad maximum in the charged particle spectrum as related to the $e^+e^-$ partial absorption energy in the Si detectors. The most impressive spectrum has been obtained with a 125-μm-thick aluminum placed in the front of a detector, which was enough to stop 3-MeV protons and detect only light charged particles [see Fig. 5(c)]. The BR values obtained for the 3-mm-thick detector are slightly higher than for the other detectors. This may be related to the uncertainty of the natural background subtraction, which is larger for thicker detectors and whose amplitude varies in the range of 10%–20%. The Geant4 calculations have likewise helped to determine absolute probabilities of high-energy bremsstrahlung emission and the neutron response function of the NaI(Tl) detector, enabling simultaneous study of all open DD reaction channels. Thus, the reaction branching ratios of the IPC channel to neutron and proton channels could be determined using different methods at two deuteron energies, 10 and 20 keV, to compare with the results obtained with the Si detectors. For a detailed description of the experimental procedure, see the Appendix. In general, absolute BR values obtained from photonic measurements are characterized by relatively large uncertainties (Table I) due to the impossibility of precisely determining the place of $\gamma$ quantum creation. In the case of neutron detection, the main problem was systematic uncertainties of cross sections of the $(n,\gamma)$ reactions in the large-volume NaI(Tl) detector. Nevertheless, the corresponding experimental spectra of all processes could be described very well. To reduce these systematic errors, a ratio between the BR values obtained at 10 and 20 keV has been determined. The obtained numbers agree very well both with the values obtained for Si detectors and with theoretical predictions.

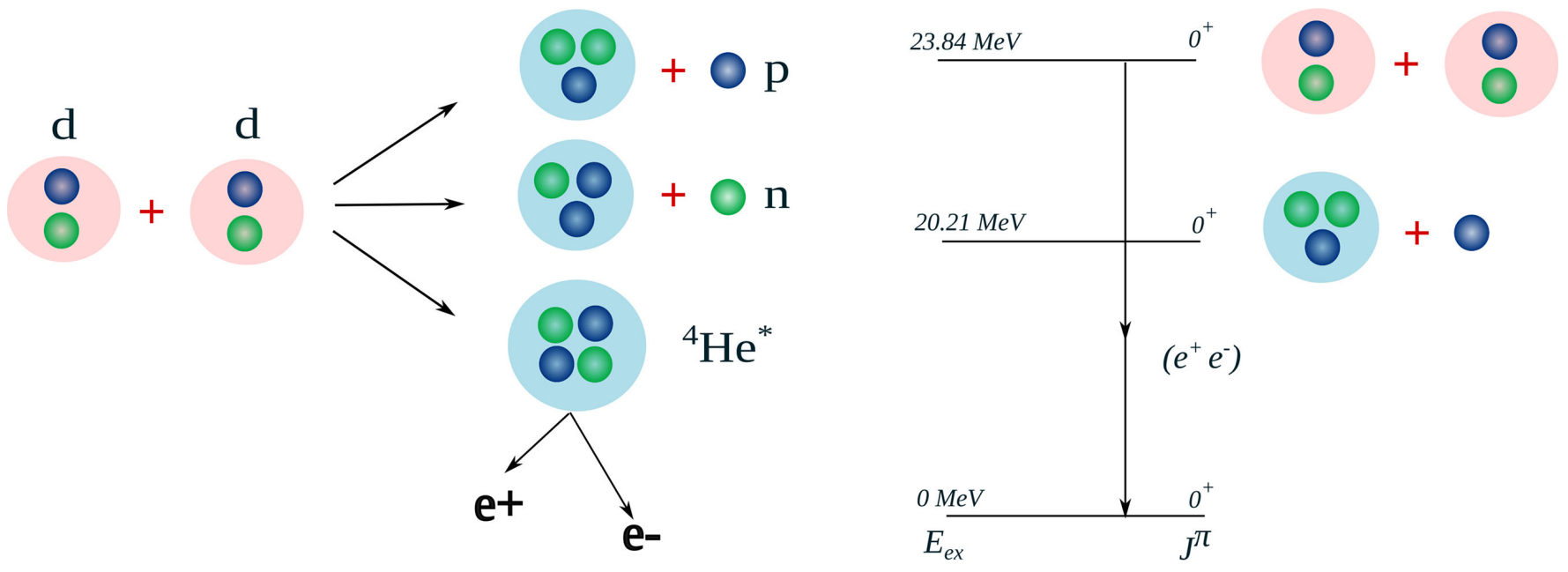

FIG. 9. Schematic visualization of the nuclear level scheme of $^4$He, showing the deexcitation of the DD threshold resonance via internal pair creations to the $0^+$ ground state.

Thus, we found clear indications of a new $e^+/e^-$ channel in DD fusion via the formation of a $^4$He threshold resonance and its decay via 22.82-MeV IPC to the ground state, as schematic represented in Fig. 9.

Improved theoretical calculations can explain the significant increase in branching ratios for lowering energies of the deuteron beam by exciting the single-particle DD threshold resonance, previously observed in the $^2\mathrm{H}(d,p)^3\mathrm{H}$ reaction [6,7]. The BR calculations have included for the first time the interference effect in both investigated proton and IPC reaction channels, which leads to a much higher partial resonance width ratio, $\Gamma_{\mathrm{IPC}}/\Gamma_p = 14$, than the one previously determined. The obtained results show that the strongest DD fusion reaction channel for deuteron energies below 5 keV is the internal pair creation $e^+e^-$, which can help in construction of future fusion energy sources if proper materials with a large electron screening energy could be utilized. The internal pair creation should also be involved in nuclear reaction rate calculations for astrophysical plasmas [4,5].

## ACKNOWLEDGMENTS

This project has received funding from the European Union's Horizon 2020 research and innovation program under Grant Agreement No. 951974.

## DATA AVAILABILITY

Some of the data that support the findings of this article are openly available [10,11,21,23,36]. The data are available upon reasonable request from the authors and are not publicly available. The data are available from the authors upon reasonable request.

## APPENDIX

### 1. Determination of branching ratios

In this appendix, the details of the procedure for determining the deuteron-deuteron fusion branching ratios between the different reaction channels $^2\mathrm{H}(d,p)^3\mathrm{H}$, $^2\mathrm{H}(d,n)^3\mathrm{He}$, and $^2\mathrm{H}(d,e^+e^-)^4\mathrm{He}$ are presented (see Table I). In accordance with the experimental data, we assumed that the branching ratio between neutron and proton channels is equal to unity with good accuracy in the studied deuteron energy region. Thus, the main attention was paid to the internal pair creation channel, for which three different detectors were used.

### 2. Si detectors

Applying Si detectors of different thicknesses, we were able to measure both proton and IPC channels simultaneously (see Fig. 10). The experimentally determined thick target yield reads as follows:

$$Y_p(E) = N_0 \int_0^R \sigma_p(E)dx = N_0 \int_0^E \frac{\sigma_p(E)}{|\frac{dE}{dx}|} dE, \quad \text{(A1)}$$

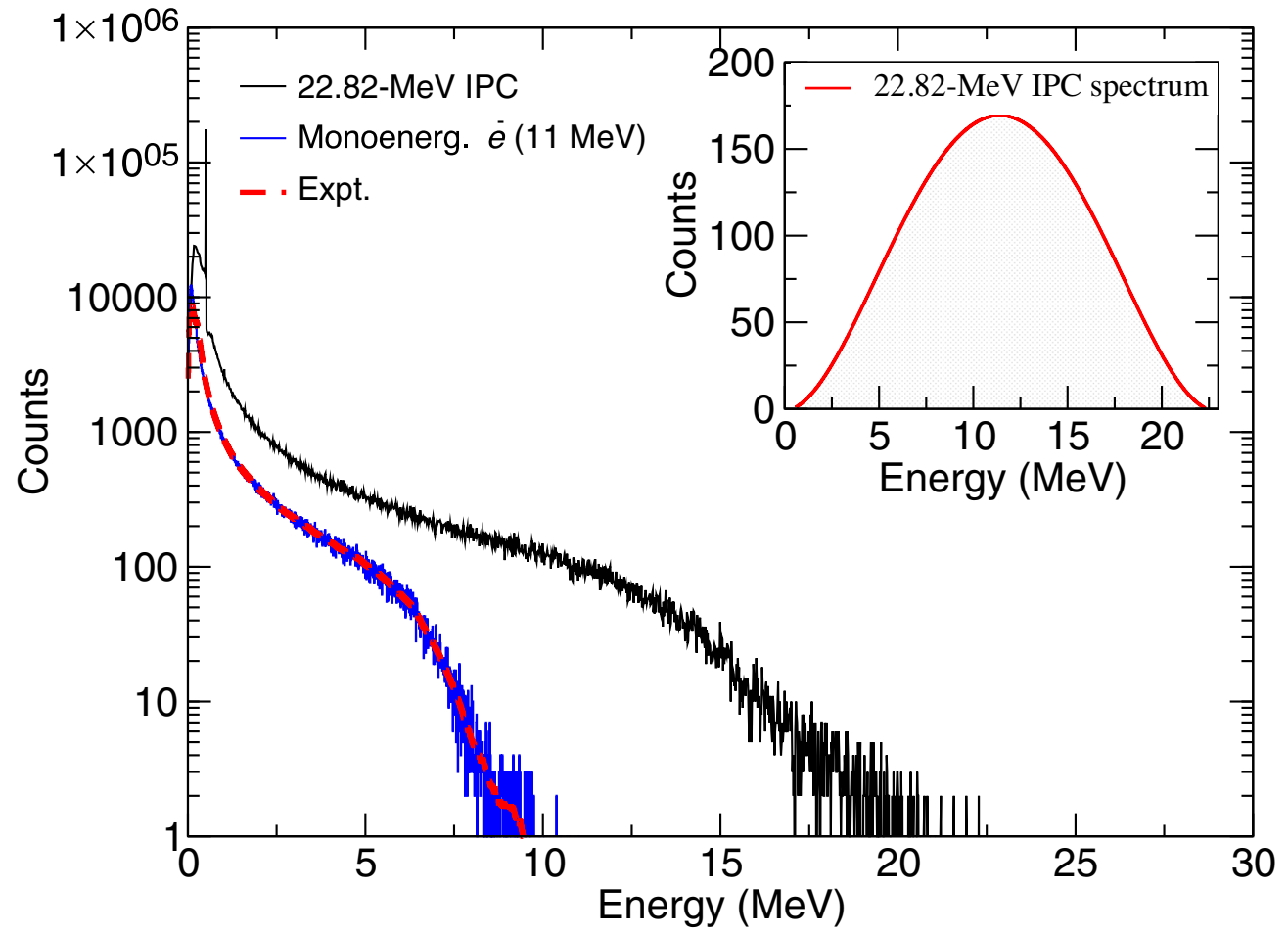

FIG. 10. The bremsstrahlung photon energy spectrum for a large-volume cylindrical NaI(Tl) detector (20 cm diameter, 30 cm length), obtained using Geant4 MC simulations (blue curve) for 11-MeV monoenergetic electrons, is shown along with the experimentally benchmarked spectrum used for validation (red dashed curve) [24,25]. Extending simulation validation, the bremsstrahlung photon spectrum simulated with Geant4 for the same size NaI detector, which induces the 22.82-MeV IPC spectrum (black curve), is also presented. The calculated 22.82-MeV IPC spectrum, resulting from the deexcitation of the DD threshold resonance via internal pair creation to the $0_0^+$ ground state, is shown in the inset.

where $N_0$ and $R$ are the density of deuterons in the target and the range of deuteron projectiles in the target, respectively. The stopping power function $dE/dx$ is proportional to $\sqrt{E}$ in the low-energy limit. We assume that the proton angular distribution is isotropic and the corresponding total cross section $\sigma_p(E)$ can be expressed by the weakly energy-dependent astrophysical $S$ factor $S_p(E)$:

$$\sigma_p(E) = \frac{1}{\sqrt{E(E+U_e)}} S_p(E) \exp\left(-\sqrt{\frac{E_G}{E+U_e}}\right), \quad \text{(A2)}$$

where the shielding of Coulomb barrier by target electrons surrounding reacting nuclei is taken into account by the screening energy $U_e$ and $E_G$ stands for the Gamow energy. The integral of Eq. (1) can be calculated analytically [37], which leads to the expression:

$$Y_p(E) = AN_0 S_p(E) \exp\left(-\sqrt{\frac{E_G}{E+U_e}}\right), \quad \text{(A3)}$$

where $A$ is an energy-independent constant. From an experimental point of view, the thick target yield can be obtained from the normalized count rate of protons $N_p^{\mathrm{exp}}$:

$$Y_p^{\mathrm{exp}}(E) = \frac{N_p^{\mathrm{exp}}}{I_d \eta_p t}, \quad \text{(A4)}$$

where $I_d$ and $t$ are the deuteron beam current and the measurement time. The detector efficiency $\eta_p = \delta \cdot \Omega_p / 4\pi$ is a product of the internal detector efficiency ($\delta = 1$ for a Si detector) and the detector solid angle. Similarly, we can determine the thick target yield of emitted electrons and positrons. Therefore, the IPC-proton BR can be finally obtained as follows:

$$\frac{Y_{\mathrm{IPC}}^{\mathrm{exp}}(E)}{Y_p^{\mathrm{exp}}(E)} = \frac{S_{\mathrm{IPC}}(E)}{S_p(E)}, \quad \text{(A5)}$$

which is independent of the screening energy and the solid angle of the detector. Hence, the only systematic uncertainty in the determination of the experimental BR is subtraction of the background. The energy dependence of the experimental branching ratio can be described by the theoretical $S$ factors.

### 3. NaI(Tl) Detector

Applying a large-volume NaI(Tl) detector, we were able to simultaneously measure the number of emitted neutrons and IPC events. The latter can be determined both from the bremsstrahlung induced by high-energy electrons or positrons and from the annihilation radiation of stopped positrons. All the effects can be studied in different energy regions of detected $\gamma$ spectra. To determine the bremsstrahlung contribution, we analyzed the high-energy spectrum region above 9 MeV. In this region, the cosmic background and the electronic noise contributions are essential to discriminate them from the bremsstrahlung components. In Fig. 4, our background measurements are presented. The spectrum shape could be very well described by the cosmic background induced by high-energy muons or electrons. The amplitude of this background is, however, subject to natural change within about 10%–20%. Furthermore, the electronic noise increases the background amplitude additionally, holding the shape of the energy spectrum. The excess of counts in this energy region compared to the background is presented in Table II and can be related to the bremsstrahlung contribution.

The experimentally determined count rate of bremsstrahlung photons can be calculated using Geant4 MC simulations, which automatically take into account the probability $w_{\mathrm{brems}}(E, E_\gamma)$ of emission of a photon with the energy $E_\gamma$ produced by an electron or positron of energy $E$:

$$N_{\mathrm{brems}}(E_\gamma) = \frac{1}{2}\int_0^E N_{\mathrm{IPC}}(E) w_{\mathrm{brems}}(E, E_\gamma) \eta_\gamma(E_\gamma) dE, \quad \text{(A6)}$$

where $N_{\mathrm{IPC}}(E)$ describes the number and energy distribution of the IPC $e^+/e^-$ and $\eta_\gamma(E_\gamma)$ stands for the detection efficiency of the NaI(Tl) detector, which was experimentally determined with help of Geant4 simulations (Fig. 3). Here, we consider that about half of electrons and positrons are stopped in the target holder, while the rest are emitted in the opposite direction and are stopped in the chamber wall much farther from the detector. This factor is not known exactly and strongly contributes to systematic uncertainty of the $N_\gamma$ value, which is of about 20%. The uncertainties related to the detector efficiency and bremsstrahlung probability are lower and are below 10%. This is validated with experimental data, where we analyzed the bremsstrahlung response function of a large-volume NaI detector to monoenergetic electrons. The results are displayed in Fig. 10. For comparison, the number of detected bremsstrahlung photons is also given for the IPC process. All the above systematic uncertainties can be significantly reduced if the DD BR will not be calculated for a single deuteron beam energy but a ratio between the values obtained for two different beam energies is determined (see Table I).

TABLE II. Experimental (Expt.) effect against background (Bkg.) for different photon energy regions between 8 and 23 MeV. In the last column, a ratio between experimental effect and background is displayed. In the 8-10-MeV energy region, there is a very small contribution of neutron-induced events [see Figs. 6(a) and 6(b)]; at energies above 10 MeV, the IPC-induced bremsstrahlung dominates.

| Energy (MeV) [1] | Total Expt. counts [2] | Bkg. counts [3] | Expt. after Bkg. Subt. counts [2-3] | Ratio [2-3]/[3] |
|---|---|---|---|---|
| 8–10 | $470 \pm 22$ | $203 \pm 14$ | $267 \pm 26$ | $1.32 \pm 0.16$ |
| 10–14 | $256 \pm 16$ | $151 \pm 13$ | $105 \pm 21$ | $0.70 \pm 0.15$ |
| 14–18 | $160 \pm 13$ | $117 \pm 11$ | $43 \pm 17$ | $0.37 \pm 0.15$ |
| 18–23 | $77 \pm 9$ | $62 \pm 8$ | $15 \pm 12$ | $0.24 \pm 0.20$ |

In the $\gamma$ energy region 4–9 MeV of the NaI detector, the energy spectrum is dominated by $\gamma$ transitions resulting from neutron capture processes within the detector scintillation crystal. After background subtraction, already estimated for higher $\gamma$ energies, the measured neutron response function could be used to determine the 2.5-MeV neutron flux arising from the DD reaction:

$$N_{n\gamma}(E_\gamma) = N_n w_{n\gamma}(E_\gamma) n_\gamma(E_\gamma), \qquad \text{(A7)}$$

where $N_{n\gamma}(E_\gamma)$ is the gamma energy spectrum induced by $N_n$ neutrons. The probability function $w_{n\gamma}(E_\gamma)$ describing the neutron capture and resulting $\gamma$'s with energy $E_\gamma$ was obtained by Geant4 MC calculations and checked using experimental data of Refs. [34,35]. However, like the bremsstrahlung analysis, there is still an uncertainty of about 10% in the absolute values of $N_{n\gamma}(E_\gamma)$ and $w_{n\gamma}(E_\gamma)$. To reduce these systematic uncertainties, we have normalized the number of neutrons $N_n$ produced at the deuteron beam energy of 20 keV to the number of protons $N_p^{\mathrm{exp}}$ observed with the Si detector [see Eq. (A4)] since the neutron-proton BR is very close to unity. In this way, the relative number of neutrons obtained at the deuteron beam energy of 10 keV could be much more exact.

To determine the number of IPC events based on the increase of the annihilation line compared to the background using NaI(Tl) and HPGe detectors, the experimental count rate was corrected for the detection efficiency $\epsilon_\gamma$, and similar to the bremsstrahlung analysis, we assumed that only half the number of positrons were stopped in the target holder and produced annihilation radiation.

[1] W. A. Fowler, *Experimental and theoretical nuclear astrophysics: The quest for the origin of the elements*, Rev. Mod. Phys. **56**, 149 (1984).
[2] C. Rolfs and W. S. Rodney, *Cauldrons in the Cosmos* (University of Chicago Press, Chicago, 1988).
[3] C. Iliadis, *Nuclear Physics of Stars* (Wiley-VCH Verlag GmbH & Co., Weinheim, 2015), Chap. 4, p. 323.
[4] Rachid Ouyed, W. R. Fundamenski, G. R. Cripps, and P. G. Sutherland, *D-D fusion in the interior of Jupiter?*, Astrophys. J. **501**, 367 (1998).
[5] S. Ichimaru and H. Kitamura, *Pycnonuclear reactions in dense astrophysical and fusion plasmas*, Phys. Plasmas **6**, 2649 (1999).
[6] K. Czerski, *DD nuclear reactions at extremely low energies*, Phys. Rev. C **106**, L011601 (2022).
[7] K. Czerski, D. Weissbach, A. I. Kilic, G. Ruprecht, A. Huke, M. Kaczmarski, N. Targosz-Ślęczka, and K. Maass, *Screening and resonance enhancements of the* $^2$H$(d,p)^3$H *reaction yield in metallic environments*, Europhys. Lett. **113**, 22001 (2016).
[8] V. A. Sergeyev, *New* $^4$He *levels and the difference in angular distributions for the reactions* D$(d,n)^3$He *and* D$(d,p)^3$H, Phys. Lett. **38B**, 286 (1972).
[9] D. R. Tilley, H. R. Weller, and G. M. Hale, *Energy levels of light nuclei* $A = 4$, Nucl. Phys. **A541**, 1 (1992).
[10] K. Czerski, R. Dubey, M. Kaczmarski, A. Kowalska, N. Targosz-Ślęczka, G. Das Haridas, and M. Valat, *Indications of electron emission from the DD threshold resonance*, Phys. Rev. C **109**, L021601 (2024).
[11] R. Dubey, K. Czerski, H. Gokul Das, M. Kaczmarski, A. Kowalska, N. Targosz-Ślęczkaczka, and M. Valat, *Observation of a new decay channel of the DD threshold resonance*, Acta Phys. Pol. B **17**, 3-A35 (2024).
[12] A. Csótó and G. M. Hale, *Nature of the first excited state of* $^4$He, Phys. Rev. C **55**, 2366 (1997).
[13] E. Hiyama, B. F. Gibson, and M. Kamimura, *Four-body calculation of the first excited state of* $^4$He *using a realistic*

*NN interaction:* $^4$He$(e,e')^4$He$(0^+)$ *and the monopole sum rule*, Phys. Rev. C **70**, 031001(R) (2004).

[14] N. Michel, W. Nazarewicz, and M. Płoszajczak, *Description of the proton-decaying* $0^+$ *resonance of the alpha particle*, Phys. Rev. Lett. **131**, 242502 (2023).

[15] H. M. Hoffmann and G. M. Hale, $^4$He *experiments can serve as a database for determining the three-nucleon force*, Phys. Rev. C **77**, 044002 (2008).

[16] K. Arai, S. Aoyama, Y. Suzuki, P. Descouvemont, and D. Baye, *Tensor force manifestations in ab initio study of the* $^2$H$(d,\gamma)^4$He, $^2$H$(d,p)^3$H, *and* $^2$H$(d,n)^3$He *reactions*, Phys. Rev. Lett. **107**, 132502 (2011).

[17] S. Aoyama, K. Arai, Y. Suzuki, P. Descouvemont, and D. Baye, *Tensor force manifestations in ab initio study of* $^2$H$(d,\gamma)^4$He, Few-Body Syst. **52**, 97 (2012).

[18] A. Cvetinović *et al.*, *Electron screening in palladium*, Phys. Lett. B **838**, 137684 (2023).

[19] A. Kowalska *et al.*, *Crystal lattice defects in deuterated Zr in the presence of O and C impurities studied by PAS and XRD for electron screening effect*, Materials **16**, 6255 (2023).

[20] M. Kaczmarski, A. I. Kilic, K. Czerski, A. Kowalska, D. Weissbach, N. Targosz-Sleczka, A. Huke, and G. Ruprecht, *New accelerator facility for measurements of nuclear reactions at energies below 1 keV*, Acta Phys. Pol. B **45**, 509 (2014).

[21] H. Gokul Das, R. Dubey, K. Czerski, M. Kaczmarski, A. Kowalska, N. Targosz-Ślęczka, and M. Valat, *High-energy electron measurements with thin Si detectors*, Measurement **228**, 114392 (2024).

[22] S. Agostinelli, Ll. Font, and C. Baixeras, *Geant4—A simulation toolkit*, Nucl. Instrum. Methods Phys. Res., Sect. A **506**, 186 (2003).

[23] R. Dubey, H. Gokul Das, K. Czerski, M. Kaczmarski, A. Kowalska, N. Targosz-Ślęczka, and M. Valat, *Energy spectra assessment of* NaI(Tl)*, NE113, and NE213 scintillators in very low-energy fusion studies*, Fusion Sci. Technol., 10.1080/15361055.2025.2520724 (2025).

[24] B. A. Faddegon, L. Van der Zwan, D. W. O. Rogers, and C. K. Ross, *Precision response estimation, energy calibration, and unfolding of spectra measured with a large* NaI *detector*, Nucl. Instrum. Methods Phys. Res., Sect. A **301**, 138 (1991).

[25] C. W. Sandifer and M. Taherzadeh, *Measurement of linac thick-target bremsstrahlung spectra using a large* NaI *scintillation spectrometer*, IEEE Trans. Nucl. Sci. **15**, 336 (1968).

[26] M. Morhac, *Deconvolution methods and their applications in the analysis of* $\gamma$*-ray spectra*, Nucl. Instrum. Methods Phys. Res., Sect. A **559**, 119 (2006).

[27] Y. Nagai, T. Kikuchi, T. Kii, T. S. Suzuki, T. Murakami, T. Shima, and T. Ohsaki, $^{127}$I$(n,\gamma)^{128}$I *reaction in* NaI(Tl) *due to cosmic-ray muon capture by Pb*, Nucl. Instrum. Methods Phys. Res., Sect. A **368**, 498 (1996).

[28] C. Hagmann, D. Lange, and D. Wright, *Cosmic-ray shower generator (CRY) for Monte Carlo transport codes*, in *Proceedings of the IEEE Nuclear Science Symposium and Medical Imaging Conference (NSS/MIC 2007): Honolulu, Hawaii,* (2007), p. 1143.

[29] Joseph W. Fowler *et al.*, *Spectroscopic measurements and models of energy deposition in the substrate of quantum circuits by natural ionizing radiation*, PRX Quantum **5**, 040323 (2024).

[30] M. B. Chadwick *et al.*, *ENDF/B-VII.0: Next generation evaluated nuclear data library for nuclear science and technology*, Nucl. Data Sheets **107**, 2931 (2006).

[31] A. I. Blokhin, E. V. Gai, A. V. Ignatyuk, I. I. Koba, V. N. Manokhin, and V. G. Pronyaev, *New version of neutron evaluated data library BROND-3.1*, Probl. At. Sci. Technol. Ser. Nucl. React. Constants **2**, 2 (2016).

[32] A. A. J. M. Plompen, O. Cabellos, C. De Saint Jean, M. Fleming, A. Algora, M. Angelone, P. Archier, E. Bauge, O. Bersillon, A. Blokhin, and F. Cantargi, *The joint evaluated fission and fusion nuclear data library, JEFF-3.3*, Eur. Phys. J. A **56**, 181 (2020).

[33] O. Iwamoto, K. Shibata, N. Iwamoto, S. Kunieda, F. Minato, A. Ichihara, and S. Nakayama, *Status of the JENDL project*, EPJ Web Conf. **146**, 02005 (2017).

[34] *Reference database for neutron activation analysis, International Atomic Energy Agency*, https://www-nds.iaea.org/naa/.

[35] K. Bergaoui, N. Reguigui, C. K. Gary, C. Brown, J. T. Cremer, J. H. Vainionpaa, and M. A. Piestrup, *Monte Carlo simulation of explosive detection system based on a deuterium–deuterium (D-D) neutron generator*, Appl. Radiat. Isot. **94**, 118 (2014).

[36] R. Dubey, K. Czerski, H. Gokul Das, M. Kaczmarski, A. Kowalska, N. Targosz-Ślęczka, and M. Valat, *Observation of electron emissions from the DD reaction: Its implications in fundamental and applied research*, Acta Phys. Pol. A **146**, 5 (2024).

[37] K. Czerski, R. Dubey, M. Kaczmarski, A. Kowalska, N. Targosz-Ślęczka, G. Das Haridas, and M. Valat, *Observation of thermal deuteron-deuteron fusion in ion tracks*, arXiv:2409.02112.